\documentclass[12pt]{article}

\usepackage{pgfplots}
\usetikzlibrary{decorations.pathmorphing}

\tikzset{snake it/.style={decorate, decoration=snake}}
\pgfplotsset{compat=1.10}
\usepgfplotslibrary{fillbetween}
\usetikzlibrary{patterns}

\usepackage{draft} 
\usepackage{hyperref}
\usepackage{graphicx,color,subfig}
\usepackage{cite}
\usepackage{mciteplus}
\usepackage{skak}
\usepackage{xcolor}
\usepackage{empheq}
\usepackage{tikz}
\DeclareFontFamily{OT1}{pzc}{}
\DeclareFontShape{OT1}{pzc}{m}{it}{<-> s * [1.10] pzcmi7t}{}
\DeclareMathAlphabet{\mathpzc}{OT1}{pzc}{m}{it}

\def\be#1\ee{\begin{align}#1\end{align}}



\begin{document}

\unitlength = .8mm

\begin{titlepage}

\begin{center}

\hfill \\
\hfill \\
\vskip 1cm

\title{\Huge On Small Black Holes in String Theory}

\author{Bruno Balthazar, Jinwei Chu,
David Kutasov}
\address{
Kadanoff Center for Theoretical Physics and Enrico Fermi Institute\\ University of Chicago, Chicago IL 60637
}
\vskip 1cm

\email{brunobalthazar@uchicago.edu, jinweichu@uchicago.edu, dkutasov@uchicago.edu}

\end{center}

\abstract{
We discuss the worldsheet sigma-model whose target space is the $d+1$ dimensional Euclidean Schwarzschild black hole. We argue that in the limit where the Hawking temperature of the black hole, $T$,  approaches the Hagedorn temperature, $T_H$, it can be described in terms of a generalized version of the Horowitz-Polchinski effective theory. For $d\ge6$, where the Horowitz-Polchinski EFT \cite{Horowitz:1997jc,Chen:2021dsw} does not have suitable solutions, the modified effective Lagrangian allows one to study the black hole CFT in an expansion in powers of $d-6$ and $T_H-T$. At $T=T_H$, the sigma model is non-trivial for all $d>6$. It exhibits an enhanced $SU(2)$ symmetry, and is described by a non-abelian Thirring model with a radially dependent coupling. The resulting picture connects naturally to the results of \cite{Soda:1993xc,Emparan:2013xia,Chen:2021emg}, that relate Schwarzschild black holes in flat spacetime at large $d$ to the two dimensional black hole. We also discuss an analogous open string system, in which the black hole is replaced by a system of two separated D-branes connected by a throat. In this system, the asymptotic separation of the branes plays the role of the inverse temperature. At the critical separation, the system is described by a Kondo-type model, which again exhibits an enhanced $SU(2)$ symmetry. At large $d$, the brane system gives rise to the hairpin brane \cite{Lukyanov:2003nj}.

}

\vfill

\end{titlepage}

\eject

\begingroup
\hypersetup{linkcolor=black}
\tableofcontents
\endgroup

\section{Introduction and summary}
\label{sec:intro}

In this note we continue our study~\cite{Balthazar:2022szl} of the Horowitz-Polchinski (HP) string/black hole transition in flat spacetime \cite{Horowitz:1996nw}.\footnote{We use many of the technical results of~\cite{Balthazar:2022szl}, but the overall picture we arrive at is different.} This transition is often discussed in Lorentzian signature, but we will focus on the Euclidean case, which is simpler, since one does not need to understand the physics beyond the horizon of the black hole, or the singularity. The Euclidean and Lorentzian problems are related, as discussed e.g. in \cite{Horowitz:1997jc}.

The problem we will address can be posed as follows. A Euclidean Schwarzschild black hole is a solution of Einstein gravity in an asymptotically flat spacetime 
${\mathbb{R}}^d\times S^1$. It is described by the metric 
\ie
\label{bh}
ds^2=f(r)d\tau^2+\frac{dr^2}{f(r)}+r^2d\Omega_{d-1}^2\ .
\fe
where $(r, \Omega_{d-1})$ are spherical coordinates on ${\mathbb{R}}^d$,
\ie
\label{fbeta}
f(r)=1-\left(\frac{r_0}{r}\right)^{d-2}~,
\fe
$r_0$ is the Schwarzschild radius, which is related to the mass of the black hole via the relation 
\ie
\label{mass}
M=\frac{(d-1)\omega_{d-1}}{16\pi G_N}r_0^{d-2}~,
\fe
$\omega_{d-1}$ is the area of the unit $(d-1)$-sphere, $G_N$ is the $d+1$ dimensional Newton constant, and $\tau$ is Euclidean time, that lives on a circle of circumference $\beta$, equal to the inverse Hawking temperature, $\beta=1/T$. It is related to the Schwarzschild radius via the relation 
\ie
\label{bbb}
\beta=\frac{4\pi r_0}{d-2}\ .
\fe
Since the background \eqref{bh}, \eqref{fbeta} is obtained by solving the classical Einstein equations, it is only valid for $r_0\gg l_p$ (the Planck scale). In weakly coupled string theory, there is a stronger constraint, since classical string theory reduces to Einstein gravity only at distances much larger than $l_s$, the string scale. Thus, the regime of validity of \eqref{bh}, \eqref{fbeta} is $r_0\gg l_s\gg l_p$, where the second inequality is due to small string coupling, $g_s\ll 1$.  

For $r_0$ of order $l_s$, the background \eqref{bh}, \eqref{fbeta} is replaced by a worldsheet conformal field theory (CFT) which asymptotes to free field theory on ${\mathbb{R}}^d\times S^1$ at large $r$, but is non-trival at finite $r$. From the point of view of this CFT, $r_0$ (or $\beta$, \eqref{bbb}) parametrizes a conformal manifold. The question is how does the CFT change when $r_0$ decreases from the classical GR regime $r_0\gg l_s$ to $r_0\sim l_s$. Of particular interest for the discussion of~\cite{Horowitz:1996nw,Horowitz:1997jc} is the nature of this CFT in the limit where $\beta$ approaches the inverse Hagedorn temperature of string theory in flat spacetime, $\beta_H$. 

In this limit, the string mode that winds once around the Euclidean time circle becomes massless \cite{Sathiapalan:1986db,Kogan:1987jd,OBrien:1987kzw,Atick:1988si}. Thus, if there is an effective field theory (EFT) description of the continuation of the solution \eqref{bh}, \eqref{fbeta} to this regime, the winding tachyon must be included in it. Moreover, the winding tachyon is known to be non-zero in the solution. This is the case already for large black holes \cite{Kutasov:2005rr,Chen:2021emg}, and is expected for small ones as well. 

A natural approach to the study of small Euclidean black holes is to write an effective action for the winding tachyon $\chi$, the radion $\varphi$, that describes the variation of the radius of the Euclidean time circle with the radial direction in ${\mathbb{R}}^d$, and other light fields, like the dilaton and the metric on ${\mathbb{R}}^d$, and look for solutions of this action that have the same symmetries and other properties as the Euclidean Black Hole (EBH). Horowitz and Polchinski (HP) wrote the leading terms in this action in \cite{Horowitz:1997jc}, and showed that for $d<6$ it has suitable solutions. 

It is natural to interpret the HP solution as the continuation of the EBH \eqref{bh}, \eqref{fbeta} to $\beta\sim \beta_H$ (but, see \cite{Chen:2021dsw} for a recent discussion of possible obstructions to this). Indeed, the two solutions have some features in common. In particular, both involve a condensate of the winding tachyon and break the $U(1)$ winding symmetry. Furthermore, both solutions have a finite classical entropy \cite{Horowitz:1997jc,Chen:2021dsw,Balthazar:2022szl}.

One problem with this interpretation is that the HP effective action does not seem to have solutions with the right properties for $d\ge 6$, while the black hole problem described above appears to make sense there.\footnote{For large $d$, the authors of \cite{Chen:2021emg} provide strong evidence that the black hole CFT exists for all $\beta\ge\beta_H$. It is natural to assume that this is the case for all $d$.} In section \ref{sec:closed} of this note, we resolve this difficulty. To do that, we treat $d$ as a continuous parameter, and focus (following our previous paper~\cite{Balthazar:2022szl}) on the region near $d=6$. We show that as $d\to 6$, the subleading terms to the ones that were kept by HP need to be retained, and when one does that, a sensible picture emerges. 

For $d=6-\epsilon$ with $0\le\epsilon\ll1$, we show that the range of temperatures in which the original HP solution is valid shrinks as $\epsilon\to 0$. The behavior of the solution beyond this region is sensitive to some subleading terms that were not included in the analysis of \cite{Horowitz:1997jc,Chen:2021dsw}. Keeping these terms allows one to analyze the solution in this regime using the EFT, in a double expansion in $\epsilon$ and $\beta-\beta_H$. As $\beta\to\beta_H$, the solution of the modified equations for all $d\le 6$ goes to zero for all $r$, as in \cite{Horowitz:1997jc}.

For $d=6+\epsilon$, we find that the effective field theory has solutions with the required properties, whose existence is again due to the presence of the subleading terms in the effective action. The modified action that we study gives the leading behavior of the solution in $\epsilon$ and $\beta-\beta_H$. To compute higher order corrections, one needs to include higher order terms in the effective action. For $\epsilon$ of order one, the appropriate language to describe the solution is the full classical string theory, i.e. the worldsheet CFT. 

For $d>6$, the solution does not go to zero as $\beta\to\beta_H$. We argue that at $\beta=\beta_H$, the corresponding CFT has an enhanced $SU(2)$ symmetry. It is described by a certain non-abelian Thirring model with an $r$ dependent coupling, that was introduced in~\cite{Balthazar:2022szl}. We comment on the relation of the resulting picture to that of Euclidean black holes at large $d$ \cite{Soda:1993xc,Emparan:2013xia,Chen:2021emg}.

The resulting picture is reminiscent of the one found for two dimensional black holes (see e.g. \cite{Witten:1991yr,Kazakov:2000pm}). The worldsheet CFT describing these black holes is exactly solvable, since it corresponds to a coset CFT, $SL(2,\mathbb{R})/U(1)$. Semiclassically, it describes a semi-infinite cigar geometry whose overall size is governed by $k$, the level of the underlying $SL(2,\mathbb{R})$ current algebra. The asymptotic radius of Euclidean time is given by $R=\sqrt kl_s$.

For large $k$ one can view the CFT as a solution of two dimensional dilaton gravity, and the stringy corrections are small. One of these corrections is a non-zero expectation value of the tachyon winding  around the Euclidean time circle. Since the radius of the circle far from the tip of the cigar is large, this tachyon is very heavy in this regime, and its profile decays rapidly at infinity. One can think of this tachyon as providing a non-perturbative (in $\alpha'$) correction to the worldsheet sigma-model.

On the other hand, as $k$ decreases, the tachyon becomes lighter, and at\footnote{In the superstring; in the bosonic string the corresponding value is $k=4$.} $k=2$, its rate of radial decay matches that of the geometric perturbation that deforms the asymptotic cylinder ${\mathbb{R}}\times S^1$ to a cigar. At that point, the $SL(2,\mathbb{R})/U(1)$ CFT develops an enhanced $SU(2)$ symmetry \cite{Murthy:2003es}. For $k<2$, the effect of the tachyon dominates over that of the geometric deformation.

The above picture is known as the FZZ correspondence \cite{FZZref,Giveon:1999px,Kazakov:2000pm,Hori:2001ax,Tong:2003ik}. It plays an important role in a number of applications of the two dimensional black hole in string theory, e.g.  \cite{Giveon:1999px,Chen:2021emg,Kazakov:2000pm}. The picture proposed in this note can be thought of as a generalization of the FZZ correspondence to Euclidean black holes in asymptotically flat spacetime.\footnote{Such a generalization was anticipated in \cite{Kutasov:2005rr}.} 

The analogy between the two cases is not perfect; for example, in the two dimensional black hole, the parameter $k$ that controls the size of the black hole also controls the central charge of the CFT, $c=3+\frac6k$, while in flat spacetime the corresponding parameter is the mass of the black hole, and the central charge is independent of it. Also, the flat spacetime analysis gives rise to the analog of the region $k\ge 2$ in the two dimensional problem; $\beta\to\beta_H$ in flat spacetime corresponds to $k\to 2$ in the $SL(2,\mathbb{R})/U(1)$ EBH. At the same time, the two systems are related via the large $d$ analysis of \cite{Soda:1993xc,Emparan:2013xia,Chen:2021emg}. The differences between them mentioned above have a natural interpretation in that context. 

In section \ref{sec:open} we discuss an open string analog of the EBH that describes two separated parallel D-branes, as one varies the distance between them~\cite{Chen:2021dsw}. For large separation, there is a solution of the DBI equations of motion where the branes are connected by a wide tube~\cite{Callan:1997kz}, and one can ask what happens to this solution as the separation between the branes decreases. 

In particular, when the distance approaches a critical one, at which a string stretched between the branes goes to zero mass, one can study the low energy effective action that includes this string. We show that this effective action has a similar structure to the closed string one, and thus the same kind of solutions. At the critical separation of the branes, there is again an enhanced $SU(2)$ symmetry (for $d>6$), this time realized in terms of a Kondo type Lagrangian (see \cite{Affleck:1995ge} and references therein), with a coupling that depends on the radial direction. 

The resulting structure is again analogous to a known solution in a two dimensional model -- the hairpin brane in a linear dilaton space \cite{Lukyanov:2003nj,Kutasov:2004dj,Nakayama:2004yx,Sahakyan:2004cq,Nakayama:2004ge}. We discuss this analogy, and the role the hairpin brane plays in the D-brane system at large $d$.

\section{Effective field theory description of a small black hole}
\label{sec:closed}

In this section, we review the structure of the Horowitz-Polchinski effective field theory (EFT), which is expected to describe small Euclidean black holes. We also discuss higher order corrections to this EFT, that are useful in some regions of parameter space.

As in section \ref{sec:intro}, we start in the Euclidean spacetime ${\mathbb{R}}^d\times S^1$, where the circumference of the Euclidean time circle,  $\beta=2\pi R$, is equal to the inverse temperature. We take the temperature to be close to the Hagedorn temperature, i.e. $\beta\simeq\beta_H=2\pi R_H$, where
\begin{equation}
	\label{tthh}
	\begin{split}
    R_H^{\rm bosonic}=2l_s\ ,\;\; R_H^{\rm type\; II}=\sqrt2l_s\ .
	\end{split}
\end{equation}
The Horowitz-Polchinski EFT is a $d$-dimensional theory obtained by reducing the classical string theory on the Euclidean time circle $S^1$, and keeping only modes that are slowly varying on ${\mathbb{R}}^d$. These modes include the metric $g_{\mu\nu}(x)$, dilaton $\phi_d(x)$, radion $\varphi(x)$, and other massless fields that will not play a role in our discussion. 

As mentioned in section \ref{sec:intro}, near the Hagedorn temperature, we also need to include in the EFT the tachyon winding once around the Euclidean time circle. Viewed as a $d$ dimensional field, $\chi(x)$, its mass is given by  
\ie
	\label{mR}
    m_\infty^2=\frac{R^2-R^2_H}{\alpha'^2}\ .
\fe
For temperatures slightly below the Hagedorn temperature, $0\le R-R_H\ll l_s$, this mass is small, $m_\infty\ll m_s=1/l_s$. $\chi$ is a complex field, whose complex conjugate corresponds to a string with the opposite orientation. The phase of $\chi$ will not play a role in our problem; thus, we will restrict to $\chi\in\mathbb{R}_+$.

In the leading approximation, the EFT only contains the fields $\varphi$, $\chi$. The other fields mentioned above  describe the back-reaction of the geometry to the non-zero $\varphi$, $\chi$, which as we argue below can be neglected in our calculations to the order that we perform them. 

The radion $\varphi$ parametrizes the local radius of the $S^1$, via the equation
\ie
\label{rx}
R(x)=R e^{\varphi(x)}~.
\fe
Here $R$ is the radius of the circle at infinity, so $\varphi\to 0$ at large $|x|$. Expanding \eqref{rx} to leading order in $\varphi$ gives rise to the action
\cite{Horowitz:1997jc,Chen:2021dsw}
	\begin{equation}
	\label{actionclcub}
	\begin{split}
I_d=\frac{\beta}{16\pi G_N}\int d^dx\left[(\nabla\varphi)^2+|\nabla \chi|^2+\left(m^2_\infty+\frac{\kappa}{\alpha'}\varphi\right)|\chi|^2\right]\ ,
	\end{split}
	\end{equation}
where 
\begin{equation}
	\label{kappabt}
	\begin{split}
    \kappa^{\rm bosonic}=8\ ,\;\; \kappa^{\rm type\; II}=4\ .
	\end{split}
	\end{equation}	
The term linear in $\varphi$ in \eqref{actionclcub} is due to the dependence of the mass of the winding tachyon on the radius \eqref{rx}. Of course, there are higher order (in $\varphi$) contributions to this mass; they will play an important role in our analysis. 	

From \eqref{actionclcub}, we find the equations of motion
	\begin{equation}
	\label{eom0}
	\begin{split}
 	\nabla^2\chi &=(m_{\infty}^2+\frac{\kappa}{\alpha'}\varphi)\chi \ ,\\
 	\nabla^2\varphi &=\frac{\kappa}{2\alpha'}|\chi|^2\ .
	\end{split}
	\end{equation}
For $m_\infty=0$, these equations have a scaling symmetry under which the fields $\chi,\varphi$ have dimension 2, and the coordinate $x$ has dimension $-1$. This symmetry can be extended to non-zero $m_\infty$ by assigning to it scaling dimension 1. This means that the mass in \eqref{actionclcub} behaves like a relevant coupling. Hence, its effects are unimportant at short distances ($|x|\ll m_\infty^{-1}$). 

For studying small black holes, we are interested in normalizable, spherically symmetric solutions of \eqref{eom0} that behave at large $r$ like\footnote{Later, we will discuss solutions of the problem with $m_\infty=0$, for which the boundary conditions at large $r$ are $\chi(r),\varphi(r)\sim1/r^{d-2}$.}
\begin{equation}
    \begin{split}
        &\chi(r)\sim r^{-\frac{d-1}{2}}e^{-m_\infty r},
        \\
        &\varphi(r)\sim r^{-d+2}.
    \end{split}
    \label{largeRbc}
\end{equation}
Such solutions exist for $d<6$, and are known as Horowitz-Polchinski solutions \cite{Horowitz:1997jc,Chen:2021dsw,Balthazar:2022szl}. They are finite at $r=0$ and monotonically approach zero at infinity, with the large $r$ behaviour given by \eqref{largeRbc}. They scale at small $m_\infty$ like $\chi,\varphi\sim m_\infty^2$, and go to zero for all $r$ as $m_\infty\to 0$.

As mentioned above, the action \eqref{actionclcub} receives contributions of higher order in $\varphi$, $\chi$, as well as contributions from other massless fields. The justification for omitting these other contributions is that they have higher scaling dimensions w.r.t. the scaling symmetry mentioned below \eqref{eom0}. 

Indeed, all the terms in the Lagrangian \eqref{actionclcub} have scaling dimension six. Terms with more powers of the fields $\varphi$, $\chi$ or more derivatives have dimensions larger than six, and thus generically can be neglected at small $m_\infty$. 

To see that the gravity fields on ${\mathbb{R}}^d$ can be neglected as well, consider the back-reaction of the HP solution on the dilaton field $\phi_d(x)$ (the analysis of the back-reaction on the metric is similar). Including the dilaton in the HP action \eqref{actionclcub} leads to an action of the form
\ie
\frac{\beta}{16\pi G_N}\int d^dxe^{-2\phi_d}\left[-4(\nabla \phi_d)^2+(\nabla\varphi)^2+|\nabla \chi|^2+m^2_\infty|\chi|^2+\cdots\right].
\label{dilatonaction}
\fe
The leading equation of motion for $\phi_d$ takes the schematic form
\ie
\nabla^2\phi_d \sim (\nabla\varphi)^2+|\nabla \chi|^2+m^2_\infty|\chi|^2+\cdots.
\fe
Thus we find that $\phi_d$ has scaling dimension four. Its leading contribution to the action \eqref{dilatonaction} is through dimension ten operators,  like $(\nabla \phi_d)^2$, $\phi_d(\nabla \varphi)^2$, etc. Hence, we can neglect them for small $m_\infty$.

As mentioned above, the Horowitz-Polchinski solutions described above only exist for $d<6$. To see why that is, it is instructive to treat $d$ as a continuous parameter, and examine these solutions in the limit $d\to 6$. This was done in \cite{Balthazar:2022szl}, where it was shown that the maximal value of $\chi$, $\varphi$, which is attained at $r=0$, grows without bound as $d\to 6$. In particular, it was shown in \cite{Balthazar:2022szl} that  
\ie
\label{chim6d}
\chi(0)\sim \frac{40\sqrt{2}\alpha'}{\kappa (6-d)}m^2_{\infty}\ .
\fe
Superficially, this seems to suggest that when $d\to 6$ with fixed $m_\infty$, the small field approximation breaks down, and the EFT 
\eqref{actionclcub} becomes unreliable. We will next argue that the actual situation is better. We can describe the region near $d=6$ in a Taylor series in $6-d$ and $m_\infty$, by including higher dimension terms in the effective Lagrangian.

\subsection{Beyond HP I: $d=6-\epsilon$}

In this and the next subsection, we study the effect of including the first subleading, dimension eight, terms in the effective Lagrangian decribed above. These terms can
be obtained by studying string scattering amplitudes. One finds (see e.g. \cite{Brustein:2021ifl})
	\begin{equation}
	\label{actioncl}
	\begin{split}
I_d=\frac{\beta}{16\pi G_N}\int d^dx\left[(\nabla\varphi)^2+|\nabla \chi|^2+\left(m^2_\infty+\frac{\kappa}{\alpha'}\varphi+\frac{\kappa}{\alpha'}\varphi^2\right)|\chi|^2+\frac{\kappa}{4\alpha'}|\chi|^4\right]\ .
	\end{split}
	\end{equation}
The $\varphi^2|\chi|^2$ term follows from the expansion of the mass of the winding tachyon, \eqref{mR}, \eqref{rx}.\footnote{This expansion also gives rise to a dimension eight term proportional to $m^2_\infty \varphi |\chi|^2$. This term can be neglected relative to the $\varphi|\chi|^2$ term in \eqref{actioncl} for $m_\infty\ll 1$.} Importantly, since the gravitational backreaction gives rise to dimension ten terms, as we discussed around eq. \eqref{dilatonaction}, we can still neglect it at this order. 

The equations of motion of the action \eqref{actioncl} are
\begin{equation}
\label{eom}
	\begin{split}
   	\nabla^2\chi &=m_\infty^2\chi+\frac{\kappa}{\alpha'}\varphi \chi+\frac{\kappa}{2\alpha'}\chi^2\chi^*+\frac{\kappa}{\alpha'} \varphi^2\chi \ ,\\
 	\nabla^2\varphi &=\frac{\kappa}{2\alpha'}|\chi|^2+\frac{\kappa}{\alpha'}|\chi|^2\varphi\ .
	\end{split}
	\end{equation}
We look for solutions satisfying the boundary conditions \eqref{largeRbc}. 

For $m_\infty=0$, eq. \eqref{eom} have the property that setting\footnote{As mentioned above, we take $\chi(r)$ to be positive, for all $r$. Thus $\varphi(r)$ is negative.}  $\chi=-\sqrt2\varphi$  is consistent with both equations, and collapses them to a single equation for $\chi$ or $\varphi$. As explained in \cite{Balthazar:2022szl}, this is due to the fact that at the Hagedorn temperature the system has an enhanced $SU(2)$ symmetry. Note that the fact that we can relate $\varphi$ and $\chi$ is sensitive to the relative coefficient of the two quartic terms in \eqref{actioncl}. Since the coefficient of $\varphi^2|\chi|^2$ is fixed by other considerations, this provides a check on the coefficient of $|\chi|^4$. 

We are interested in the behavior of the solutions of \eqref{eom} for $\epsilon=6-d\ll 1$ and $m_\infty\ll 1$ (in string units). We next show that in this region of parameter space, we can find a relation between $\chi(0)$, which sets the scale of the solution, and the parameters $m_\infty$ and $\epsilon$. 

To this end, in Appendix \ref{app:EFTscaling}, we show that a saddle point of the action \eqref{actioncl} satisfies the relation \eqref{mchiint}, which we rewrite here for convenience: 
\begin{equation}
    \begin{split}
      m_{\infty}^2\int d^d x  |\chi_*|^2=\frac{1}{4}\int d^dx\left[(d-6)\frac{\kappa}{\alpha'}\varphi_*|\chi_*|^2+(2d-8)\left(\frac{\kappa}{\alpha'}\varphi_*^2|\chi_*|^2+\frac{\kappa}{4\alpha'}|\chi_*|^4\right)\right]\ .
    \end{split}
    \label{actionscaling}
\end{equation}
Here, $\varphi_*(x)$ and $\chi_*(x)$ are solutions to the equations of motion \eqref{eom}, satisfying the boundary conditions \eqref{largeRbc}.

As mentioned above, we are interested in the properties of the solutions for small $\epsilon$ and $m_\infty$. We can parametrize the small $\epsilon$ region by writing 
\ie
m_\infty=\epsilon y\ ,
\label{mscale}
\fe
and considering the limit $\epsilon\to 0$ with $y$ held fixed (i.e. letting $m_\infty$ scale like $\epsilon$). Assuming that the integrals in \eqref{actionscaling} are analytic in $\epsilon$ in this limit, we can compute them by perturbing around the point $m_\infty=\epsilon=0$. Thus, we are looking for normalizable solutions to the equations
\begin{equation}
\label{eom1}
	\begin{split}
   	{\nabla}^2{\chi} &=\frac{\kappa}{\alpha'}{\varphi} {\chi} \ ,\\
 	{\nabla}^2{\varphi} &=\frac{\kappa}{2\alpha'}{\chi}^2\ ,
	\end{split}
	\end{equation}
in $d=6$. In \eqref{eom1}, we assumed that $\chi(0)\ll1$, so that the cubic terms on the r.h.s of \eqref{eom} can be ignored. We will justify this assumption shortly.

As shown in~\cite{Balthazar:2022szl}, eq. \eqref{eom1} has a family of solutions with the required boundary conditions, labeled by $\chi(0)$. These solutions have  $\chi=-\sqrt{2}\varphi$ for all $r$, and
\ie
\chi(r)=\frac{\chi(0)}{\left(1+\frac{\kappa}{24\sqrt{2}\alpha'}\chi(0)r^2\right)^2}\ ~.
\label{d6m0sol}
\fe
For small $\epsilon$, we can evaluate the integrals in \eqref{actionscaling} by substituting the solution \eqref{d6m0sol} for $\chi(r)$. This leads to
\ie
\label{mchi0eq}
m_\infty^2=\frac{\sqrt{2}\kappa}{80\alpha'}(6-d)\chi(0)+\frac{3\kappa}{140\alpha'}\chi^2(0)\ .
\fe
Note that the quadratic term in $\chi(0)$ on the r.h.s. of \eqref{mchi0eq} is due to the quartic terms in the action \eqref{actioncl}, and is thus absent in the corresponding analysis for the HP action \eqref{actionclcub}. Thus, in that case, eq. \eqref{mchi0eq} reduces to \eqref{chim6d}. 

Equation \eqref{mchi0eq} implies that in the limit $\epsilon\to 0$ with fixed $y$ \eqref{mscale}, $\chi(0)$ behaves as 
\ie
\chi(0)=\epsilon F(y)+O(\epsilon^2)\ .
\label{chioo}
\fe
The function $F(y)$ can be obtained by substituting \eqref{mscale}, \eqref{chioo} into \eqref{mchi0eq}. This leads to 
\ie
\label{ffyy}
y^2=\frac{\sqrt{2}\kappa}{80\alpha'}F(y)+\frac{3\kappa}{140\alpha'}F^2(y)\ .
\fe
Thus, for $y\ll 1$ (i.e. $m_\infty\ll\epsilon$), $F$ behaves like 
\ie
\label{smally}
F(y)\sim\frac{80\alpha'}{\sqrt{2}\kappa}y^2\ ,
\fe
which is equivalent to \eqref{chim6d}, while for $y\gg 1$ (or $m_\infty\gg\epsilon$) it behaves like
\ie
F(y)\sim\sqrt{\frac{140}{3}\frac{\alpha' }{\kappa}}y \ ,
\label{dless6chivsm}
\fe
which implies, via \eqref{mscale}, \eqref{chioo}, that 
\ie
\chi(0)\sim\sqrt{\frac{140}{3}\frac{\alpha' }{\kappa}}m_\infty \ .
\label{dless6chivsm2}
\fe
In both cases, we can trust the effective action \eqref{actioncl}, as long as we restrict to $m_\infty\ll1$.

\begin{figure}[h]
\centering
 \includegraphics[width=0.5\textwidth]{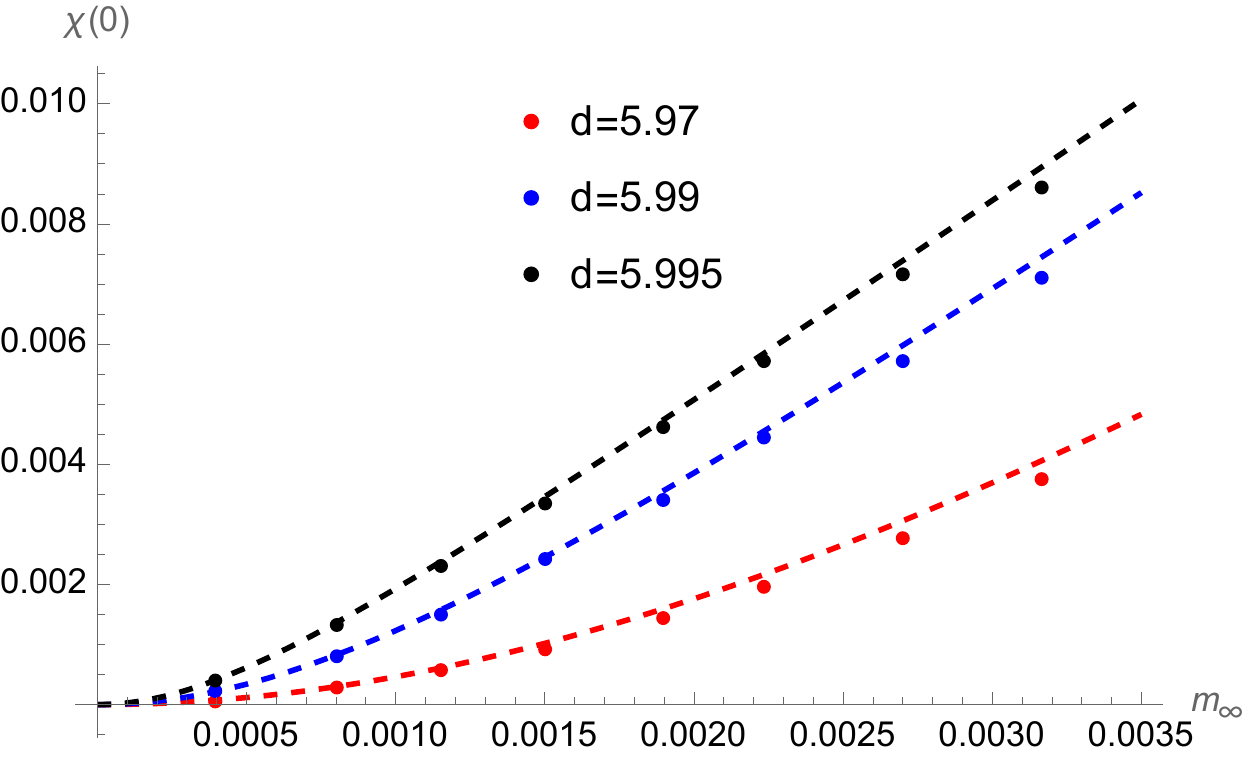}
 \caption{$\chi(0)$ as a function of $m_\infty$, with $\frac{\kappa}{\alpha'}=4$. The dashed lines are given by \eqref{mchi0eq}.} 
\label{chi0mcl}
\end{figure}

A few comments are in order at this point:
\begin{enumerate}
\item In going from \eqref{actionclcub} to \eqref{actioncl} we added the first subleading, dimension eight, terms, but neglected all the terms of higher dimension. This is justified by the results. Indeed, higher dimension terms would contribute terms of higher order in $\chi(0)$ on the r.h.s. of \eqref{mchi0eq}. Since the coefficients of these terms are not expected to have any singularities as $\epsilon, m_\infty\to 0$, their contributions to the analysis above are subleading in $m_\infty$, $\epsilon$.
    
\item In going from \eqref{eom} to \eqref{eom1} we neglected the cubic terms on the r.h.s. This is justified by the results, since $\chi(r)\ll 1$ for all $r$ for the solutions. 
    
\item To derive eq. \eqref{mchi0eq}, we substituted into the integrals \eqref{actionscaling} the $\epsilon=m_\infty=0$ solution \eqref{d6m0sol}, and not the solution for the appropriate value of $\epsilon$, $m_\infty$. Assuming the smoothness in $\epsilon$ mentioned above, the corrections to \eqref{mchi0eq} due to this approximation are subleading in these parameters.
    
\item To check eq. \eqref{mchi0eq}, we solved eq. \eqref{eom} numerically. In figure \ref{chi0mcl}, we plot $\chi(0)$ as a function of $m_\infty$, for $d$ slightly below six. We see that the agreement of the numerical results with the analytic formula \eqref{mchi0eq} improves as $6-d$ and $m_\infty$ decrease.
    
\end{enumerate}

\subsection{Beyond HP II: $d=6+\epsilon$}

In the previous subsection we saw that as $d$ approaches six from below, the region of temperatures close to the Hagedorn temperature for which the original HP analysis is valid becomes smaller and smaller. For fixed temperature, in this limit the solution of the EFT becomes more and more sensitive to the higher order terms in the effective action. 

This makes it interesting to ask what happens for $d\ge 6$. Recall that in the absence of the quartic terms in eq. \eqref{eom}, there are no normalizable solutions for $d\ge 6$, but as we discuss next, this changes when we add these terms. 

The case $d=6$ can be viewed as the limit $\epsilon\to 0$ of the analysis in the previous subsection. The solution behaves like $\chi(0)\sim m_\infty+O\left(m_\infty^2\right)$, as in \eqref{dless6chivsm2}. Its existence relies on the presence of the quartic terms in \eqref{actioncl}. Adding higher dimension terms to the effective Lagrangian gives contributions to $\chi(0)$ of higher order in $m_\infty$. 

\begin{figure}[h!]
\centering
 {\subfloat{\includegraphics[width=0.48\textwidth]{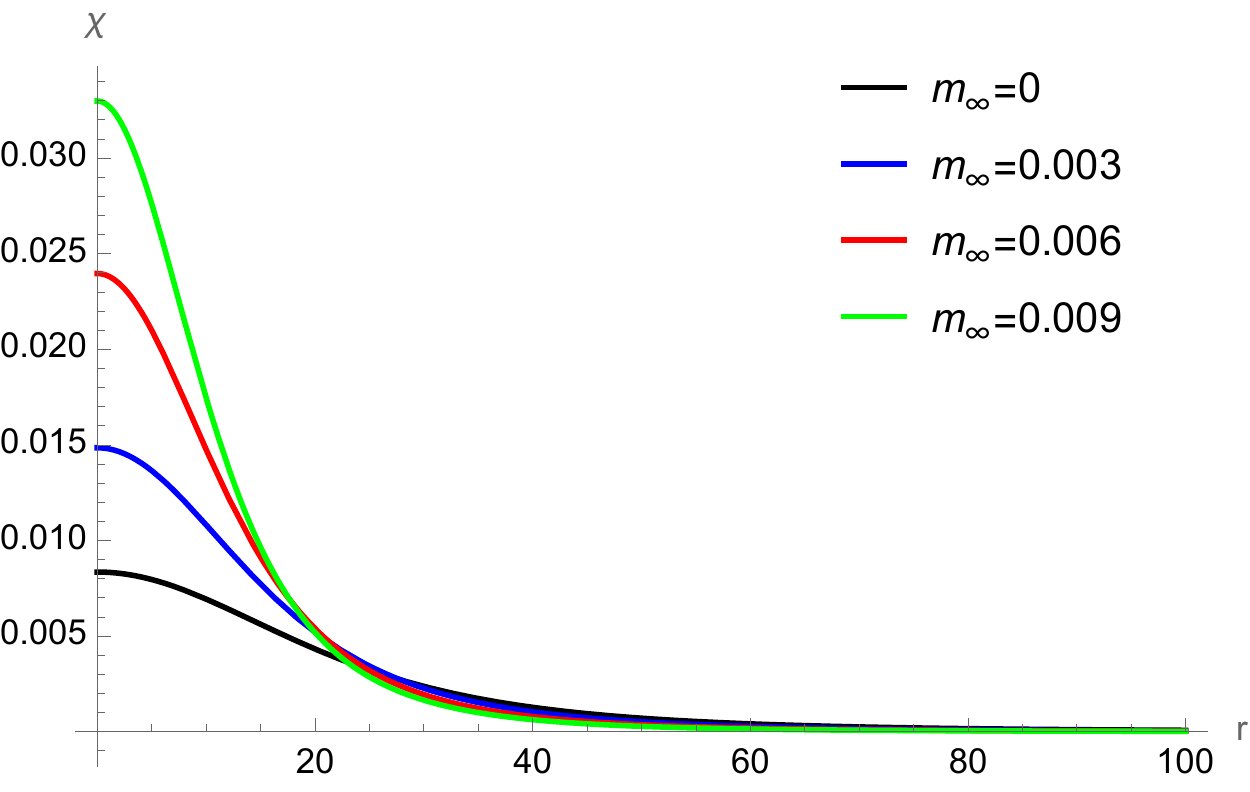}\label{chifig}}}~ 
 {\subfloat{\includegraphics[width=0.48\textwidth]{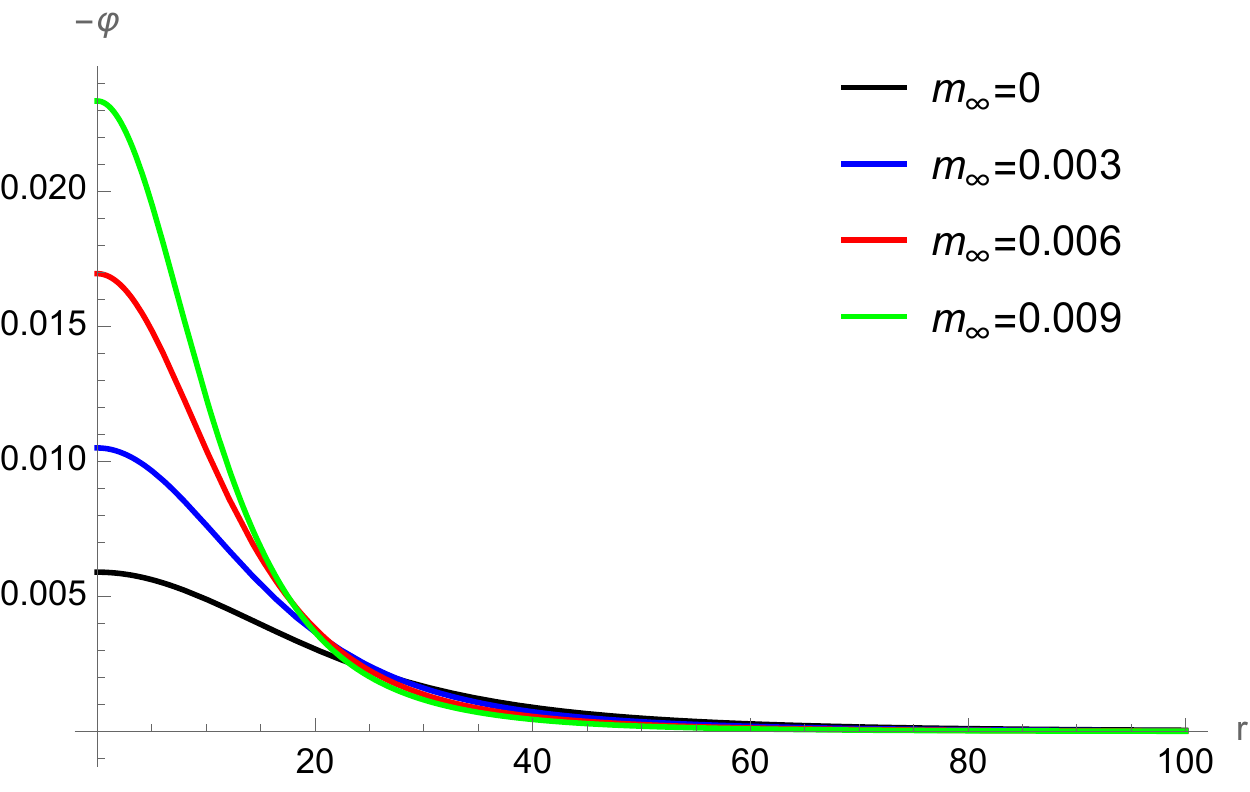}\label{phifig}}} 
 \caption{The profiles of $\chi$ and $-\varphi$ for $d=6.01$.}
 \label{chiphifigs}
\end{figure}

To see what happens for $d=6+\epsilon$, we look back at eq. \eqref{mchi0eq}. One interesting feature is that there is now a solution to this equation for $m_\infty=0$. It is
\ie
\label{chi0dg6}
\chi(0)=\frac{7\sqrt{2}}{12} \epsilon+O\left(\epsilon^2\right)\ .
\fe
This is different from the situation for $d<6$, where the HP solutions vanish at $m_\infty=0$, even after including the higher dimension terms to the effective Lagrangian. To verify that such solutions indeed exist, we solved eq. \eqref{eom} numerically. The results are exhibited in figure \ref{chiphifigs}. In this figure we also plot the solutions for $m_\infty>0$. The qualitative form of these solutions is similar to that of the HP solutions in $d<6$: they are finite at $r=0$, and monotonically approach zero at infinity, with the large $r$ behaviour \eqref{largeRbc}.

\begin{figure}[h]
\centering
 \includegraphics[width=0.5\textwidth]{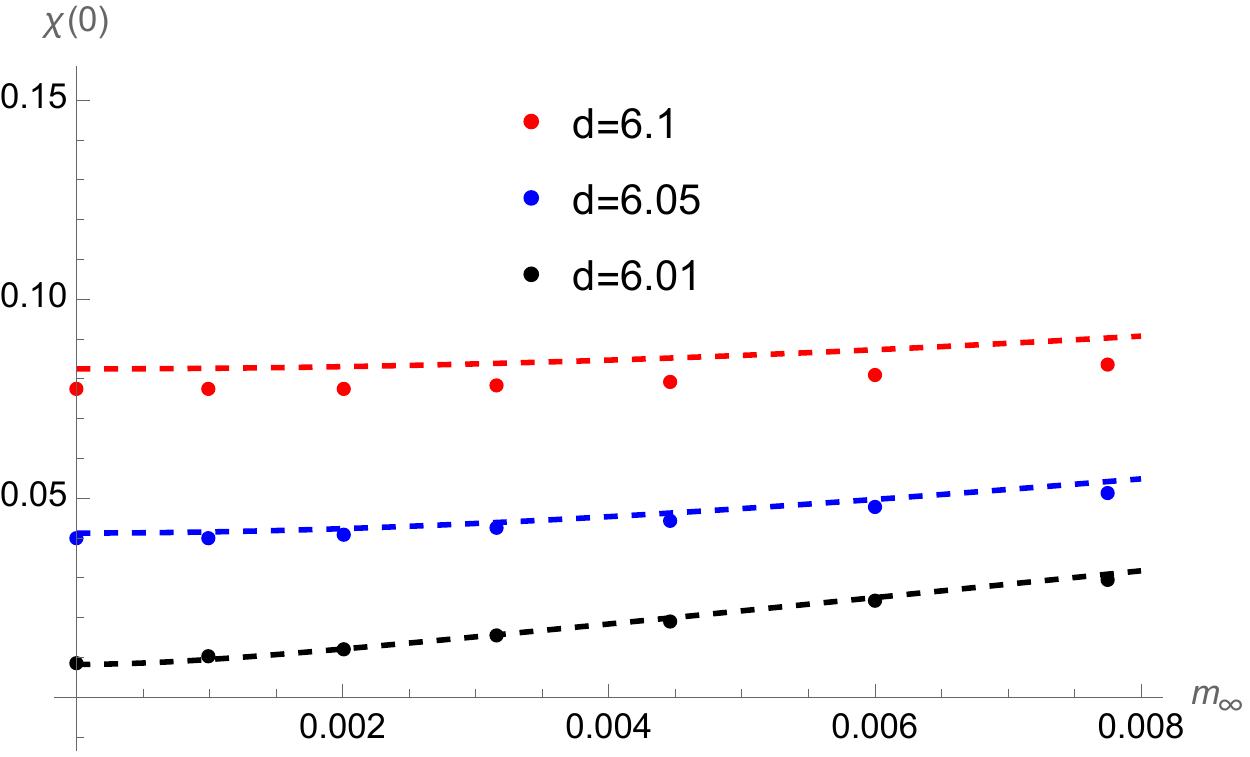}
 \caption{$\chi(0)$ as a function of $m_\infty$ for $d>6$, with $\frac{\kappa}{\alpha'}=4$. The dashed lines are given by \eqref{mchi0eq}.} 
\label{chi0mcldg6}
\end{figure}

In figure \ref{chi0mcldg6}, we plot the value of $\chi(0)$ as a function of $m_\infty$ and compare it with \eqref{mchi0eq}. We see that for small $\epsilon(=d-6)$ and $m_\infty$, \eqref{mchi0eq} is consistent with the numerical results, as expected.

Comments:
\begin{figure}[h]
\centering
 \includegraphics[width=0.5\textwidth]{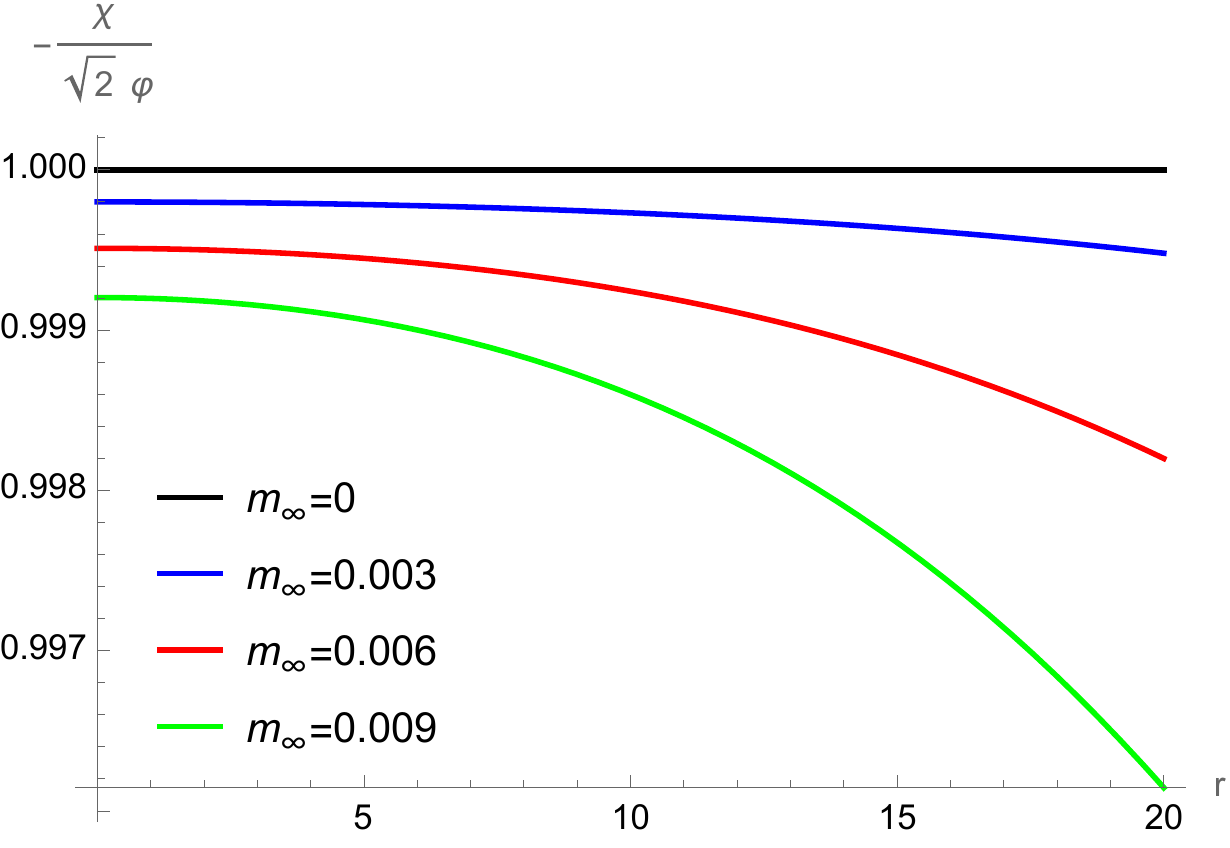}
 \caption{Plots of $-\frac{\chi(r)}{\sqrt{2}\varphi(r)}$ for different values of $m_\infty$, at $d=6.01$.} 
\label{phiovchi}
\end{figure}
\begin{enumerate}

\item For $m_\infty=0$, we find numerically that the solution of \eqref{eom} satisfies the constraint $\chi(r)=-\sqrt2\varphi(r)$ (see figure \ref{phiovchi}). 
As explained in~\cite{Balthazar:2022szl}, the constraint $\chi(r)=-\sqrt2\varphi(r)$ is a consequence of an $SU(2)$ symmetry of the underlying worldsheet CFT. This is compatible with the expectations expressed in that paper that at the Hagedorn temperature, the small EBH should exhibit such an enhanced symmetry.  

\item For $d$ slightly above six and small $m_\infty$, the EFT analysis based on the action \eqref{actioncl} is reliable. Eq. \eqref{chi0dg6} gives the order $\epsilon$ contribution to $\chi(0)$ at $m_\infty=0$. To compute higher order contributions, one needs to add higher order terms to the effective action (and include the gravitational back-reaction discussed around eq. \eqref{dilatonaction}). The coefficient of a particular power of $\epsilon$ in \eqref{chi0dg6} is only sensitive to terms up to a particular dimension in the effective action.

\item As mentioned above, for $m_\infty=0$ the worldsheet CFT describing the small EBH has an enhanced $SU(2)$ symmetry (for all $d>6$). As discussed in~\cite{Balthazar:2022szl}, it corresponds to a deformation of the flat space sigma model on $\mathbb{R}^d\times S^1$ by the non-abelian Thirring deformation
\ie
\chi(r) J^a \bar J^a,
\label{JJbar}
\fe
where $J^a$ and $\bar J^a$ are $SU(2)_L$ and $SU(2)_R$ currents that exist when the $S^1$ is at the Hagedorn radius. The radial profile $\chi(r)$ is fixed by the requirement that the corresponding worldsheet theory is conformal.

\item While the relative sign of the two quartic terms in \eqref{actioncl} is fixed by the $SU(2)$ symmetry discussed above, to compute their overall sign one needs to appeal to a string theory calculation.\footnote{Of course, another check on that sign comes from the expansion of the radius \eqref{rx} mentioned above.} As a check on that calculation, it is interesting to ask what would happen if the sign was opposite. Looking back at \eqref{mchi0eq}, one finds that in that case there would be no solution to the equations for $d\ge 6$, while for $d<6$ there would be no solution above some maximal value of $m_\infty$, and two solutions below this value. All this would be hard to interpret from the point of view of this note, which strongly supports the sign in \eqref{actioncl}. 

\end{enumerate}

The authors of~\cite{Soda:1993xc,Emparan:2013xia,Chen:2021emg} showed that the EBH geometry \eqref{bh}, \eqref{fbeta} has the interesting property that as $d\to\infty$ its reduction to the two dimensional space labeled by $(r,\tau)$ describes the same background as the $SL(2,\mathbb{R})/U(1)$ coset~\cite{Witten:1991yr}. The level of $SL(2,\mathbb{R})$, $k$, is related to the temperature via the relation $\beta=2\pi\sqrt{k}l_s$. The original papers~\cite{Soda:1993xc,Emparan:2013xia}  considered the case where the black hole is large, i.e. the inverse Hawking temperature $\beta\gg l_s$, or $k\gg1$. Their discussion has been extended to $\beta\sim l_s$ in~\cite{Chen:2021emg}, and it is natural to compare their results to ours.\footnote{In weakly coupled string theory the space dimension $d$ is bounded from above, $d\le 9$. Thus, one may think that the large $d$ limit does not make sense in this context. However, as discussed in~\cite{Balthazar:2022szl} and in section \ref{sec:intro}, one can study the EBH sigma model as a worldsheet CFT for any $d$, and in this context the limit $d\to\infty$ is sensible.}

One can view the backgrounds described in~\cite{Soda:1993xc,Emparan:2013xia,Chen:2021emg} and here as points in the two dimensional plane labeled by $(m_\infty, d)$. The work of~\cite{Soda:1993xc,Emparan:2013xia,Chen:2021emg} concerns a slice of this space of theories corresponding to large $d$ and any $m_\infty$, while our discussion is restricted to small $m_\infty$ and (in principle, if we can solve the deformed CFT \eqref{JJbar}) any $d$. The region of overlap of the two is small $m_\infty$ and large $d$. 

The simplest case to compare is $m_\infty=0$. Our proposal is that it is described by the CFT generated by the non-abelian Thirring deformation \eqref{JJbar}, and in particular has an enhanced $SU(2)$ symmetry for all $d>6$. According to~\cite{Chen:2021emg}, at large $d$ this background is described by a level two (in the superstring), or four (in the bosonic string) $SL(2,\mathbb{R})/U(1)$ coset CFT, which is known to have such an enhanced symmetry\footnote{More precisely, this is known for the worldsheet supersymmetric case, and is believed to also be the case for the bosonic one.} \cite{Murthy:2003es}. Thus, the two pictures are compatible in the overlap region. 

From our perspective, the claim of~\cite{Chen:2021emg} is that as $d\to\infty$, the non-abelian Thirring model \eqref{JJbar} approaches (after integrating out the angular d.o.f.) the $SL(2,\mathbb{R})/U(1)$ coset CFT with $k=2$. This is a region where the EFT we studied fails, since $\chi(0)$ \eqref{chi0dg6} is large, and the non-abelian Thirring model \eqref{JJbar} is strongly coupled. However, it seems quite reasonable that this indeed happens, since the $SL(2,\mathbb{R})/U(1)$ coset with $k=2$ also has a description as a non-abelian Thirring model with a coupling that depends on the radial coordinate \cite{Murthy:2003es}. More generally, the relation between our results and those of~\cite{Soda:1993xc,Emparan:2013xia,Chen:2021emg} is a kind of duality, where their description is good at large $d$ whereas ours is good near $d=6$.  

\subsection{Relation to large Euclidean black holes}

The ADM construction relates the mass of a gravitational solution to the behavior of $g_{00}$ at large $r$~\cite{Arnowitt:1962hi}. In our parametrization, if $\varphi$ behaves at large $r$ like  
\ie
\label{lr}
\begin{split}
\varphi(r)&\sim -\frac{C_\varphi}{r^{d-2}}\ ,
\end{split}
\fe
the ADM mass is given by 
\begin{equation}
\label{massHP}
\begin{split}
M=2(d-2)\omega_{d-1}\frac{C_\varphi}{16\pi G_N}\ .
\end{split}
\end{equation}
For large black holes, $C_\varphi$ can be read off \eqref{bh} -- \eqref{bbb}, 
\ie
\label{largebh}
C_\varphi=\frac{d-1}{2(d-2)}\left(\frac{(d-2)\beta}{4\pi}\right)^{d-2}\ .
\fe
If, as we proposed, as $r_0$ decreases, and in particular $\beta$ \eqref{bbb} approaches the Hagedorn temperature \eqref{tthh}, the EBH solution morphs into a solution of the (modified) HP action \eqref{actioncl}, we can compute the dependence of $C_\varphi$ on $\beta$ in that regime as well. 

Integrating the second equation of \eqref{eom}, we have
\ie
\varphi(r)\sim-\frac{\kappa}{2(d-2)\omega_{d-1}\alpha'}\int d^dx\frac{\chi^2(x)}{|\vec{r}-\vec{x}|^{d-2}}\ ,
\fe
where we have neglected higher order terms in $\chi,\varphi$. For large $r$, $\varphi$ has the behavior \eqref{lr}, with
\ie
\label{cphi}
C_\varphi\sim\frac{\kappa}{2(d-2)\omega_{d-1}\alpha'}\int d^dx\chi^2(x)\ .
\fe
As before, for $d$ close to six, and $\beta$ close to $\beta_H$, we can approximate the solution by \eqref{d6m0sol}. Plugging it in \eqref{cphi}, we get the relation between $C_\varphi$ and $\chi(0)$. 
\ie
\label{smallbh}
C_\varphi\sim\frac{\alpha'^2}{\kappa^2}\frac{576\sqrt{2}}{\chi(0)}\ .
\fe
Here, $\chi(0)$ is a solution of \eqref{mchi0eq}. In particular, for $d<6$, it goes to zero as $\beta\to \beta_H$, so $C_\varphi$ diverges in this limit. For $d>6$, $\chi(0)$ approaches a finite value of order $d-6$ as $\beta\to \beta_H$ \eqref{chi0dg6}. Therefore, $C_\varphi\sim \frac{1}{d-6} $ remains finite. 

Thus, we can compute the ADM mass \eqref{massHP} both for large black holes, with $\beta\gg l_s$, \eqref{largebh}, and for small ones, with $\beta\sim\beta_H$, \eqref{smallbh}. The interpolating behavior remains an open problem.

\section{Open string analog}
\label{sec:open}

As pointed out in \cite{Chen:2021dsw}, there is an interesting analog of the EBH problem discussed in section \ref{sec:closed}, that involves open strings rather than closed ones. To describe it, we start with two D-branes in the 
bosonic string, or a D-brane and an anti D-brane in the superstring, extended in $d$ of the $d+1$ dimensions in   
${\mathbb{R}}^{d+1}$. We will denote the $d$ directions along the branes by $x$, and the  transverse direction by $X$. The branes are initially separated by the distance $L$ in $X$; e.g., we can place them at $X=\pm L/2$. $L$ plays a similar role in the open string system to that played by the inverse temperature $\beta$ in the closed string case discussed in section \ref{sec:closed}.

It was shown in \cite{Callan:1997kz} that for large $L$ the open string equations of motion have a solution that describes the two branes connected by a wide tube. This solution can be obtained from the DBI action for the branes 
\ie
\label{dbiact}
I_\text{DBI}=\tau_d\int d^dx\sqrt{1+(\vec\nabla X)^2}\ ,
\fe
where $\tau_d$ is proportional to the tension of the branes, and $X(x)$ is the local distance between the branes. For spherically symmetric configurations, the action \eqref{dbiact} takes the form
\ie
I_\text{DBI}=\tau_d\omega_{d-1}\int dr\ r^{d-1}\sqrt{1+(X'(r))^2}\ .
\fe
A solution of the e.o.m. of this action found in \cite{Callan:1997kz} is
\ie
\label{lphi}
X(r)=\pm \left(\frac{L}{2}-\int_r^\infty dr' \frac{1}{\sqrt{\left(r'/r_\text{min}\right)^{2d-2}-1}}\right)\ .
\fe
As $r\to\infty$, \eqref{lphi} describes two branes located at $X=\pm L/2$. As $r$ decreases, the two branes get closer, and eventually, at $r=r_{\rm min}$, they meet, at $X=0$. Thus, it describes the two branes connected by a throat of width $2r_{\rm min}$.

Setting $r=r_{\rm min}$ in \eqref{lphi} gives a relation between $r_{\rm min}$ and $L$, 
\ie
\label{Lrmin}
L=2\int_{r_\text{min}}^\infty dr' \frac{1}{\sqrt{\left(r'/r_\text{min}\right)^{2d-2}-1}}
=\sqrt{\pi}\frac{d}{d-1}\frac{\Gamma \left(\frac{d}{2-2d}\right)}{\Gamma \left(\frac{1}{2-2d}\right)}r_\text{min}\ .
\fe
The DBI analysis is valid for $L\gg l_s$, where the shape of the brane \eqref{lphi}, \eqref{Lrmin} is slowly varying on the string scale. One can think of the solution in this limit as an analog of a large EBH. 

Similarly to the closed string case, the above solution contains a non-zero condensate of the open string tachyon stretched between the two branes, $\chi_\text{op}$. For large $L$, this condensate can be computed by evaluating the area of the worldsheet of a string ending on the curved D-brane \eqref{lphi}, \cite{Aharony:2008an}. Like in the closed string case, this tachyon is very heavy at large $r$, and therefore its profile decays rapidly there. As there, it gives a non-perturbative $\alpha'$ correction to the EFT (DBI) analysis. 

When the asymptotic distance between the two branes decreases, the width of the tube connecting them shrinks, \eqref{Lrmin}. Like in section \ref{sec:closed}, it is natural to ask what happens to the solution when $L$ approaches the string scale, and in particular when the asymptotic mass of the open string tachyon $\chi_\text{op}$ goes to zero.

The same logic as in section \ref{sec:closed} leads us to consider an EFT containing the tachyon $\chi_\text{op}$, and the field parametrizing the local variation of the distance between the D-branes from $L$, $\varphi_\text{op}$. In Appendix \ref{Smat} we study this action, to the same order as in the closed string case \eqref{actioncl}. We do the calculation for the bosonic string; we expect the qualitative structure to be the same for the superstring, but have not checked this. 

The resulting action is given by
\ie
\label{opeft}
I_\text{op}=\frac{\tau_dL^2}{2}\int d^dx\left[(\nabla\varphi_\text{op})^2+|\nabla \chi_\text{op}|^2+\left(m^2_\infty+\frac{\kappa}{2\alpha'}\varphi_\text{op}+\frac{\kappa}{2\alpha'}\varphi_\text{op}^2\right)|\chi_\text{op}|^2+\frac{\kappa}{8\alpha'}|\chi_\text{op}|^4\right]\ ,
\fe
where $\kappa$ is given by \eqref{kappabt}. As before, $m_\infty$ is the asymptotic mass of the stretched tachyon. It is given by  
\begin{equation}
\label{moL}
	\begin{split}
 m_{\infty}^2=\frac{L^2-L_c^2}{(2\pi \alpha')^2}\ .
	\end{split}
	\end{equation}
where
\begin{equation}
\label{llcc}
	\begin{split}
  L_c^{\rm bosonic}=2\pi l_s\ ,\;\; L_c^{\rm type\; II}=\sqrt2 \pi l_s\ .
	\end{split}
	\end{equation}
As in section \ref{sec:closed}, we expect \eqref{opeft} to be valid for $m_\infty\ll m_s$, i.e. for $L$ close to the critical value \eqref{llcc}. 

We are again interested in normalizable, spherically symmetric solutions of the e.o.m. of \eqref{opeft}. Since the action is  the same as in the closed string case, \eqref{actioncl}, so are the solutions. In particular, for $d>6$, $m_\infty=0$, the solutions satisfy 
\ie
\label{chieqphi}
\chi_\text{op}(r)=-\sqrt{2}\varphi_\text{op}(r)\ .
\fe
In the closed string case, this relation had a natural worldsheet interpretation, stemming from an enhanced $SU(2)$ symmetry of the non-abelian Thirring model \eqref{JJbar}. It is natural to expect that something similar happens in the open string case. 

To see that this is indeed the case, we consider the worldsheet action for the system of two D-branes in the presence of the perturbations corresponding to $\chi_\text{op}$ and $\varphi_\text{op}$. Keeping track of normalizations, the free worldsheet Lagrangian for the two branes contains in this case the boundary interaction
\begin{equation}
\label{delcurr}
	\begin{split}
 	\delta L=-\varphi_\text{op}(r)J^3\sigma^3+\frac{1}{\sqrt2}\chi_\text{op}(r)J^+\sigma^-+\frac{1}{\sqrt2}\chi_\text{op}^*(r)J^-\sigma^+\ .
	\end{split}
	\end{equation}
$J^i$ are given by \eqref{Jplus}, \eqref{J3}, and $\sigma^i$ are Pauli matrices (with $\sigma^{\pm}\equiv (\sigma^1\pm i\sigma^2)/2$), whose role is to keep track of the Chan-Paton structure associated with the two branes.

When \eqref{chieqphi} is satisfied, we can write \eqref{delcurr} as
\begin{equation}
\label{SU2delcurr}
	\begin{split}
 	\delta L=\chi_\text{op}(r)J^i\sigma^i\ .
	\end{split}
	\end{equation}
This boundary interaction is reminiscent of the boundary CFT	description of the Kondo effect~\cite{Affleck:1995ge}. In that case, the ${\mathbb{R}}^d$ is absent, and the coupling \eqref{SU2delcurr} generates an RG flow. The infrared fixed point of that flow describes a system where the $SU(2)$ global symmetries corresponding to $J^i$, $\sigma^i$ are broken, but the total $SU(2)$ corresponding to $J^i+\sigma^i$ remains a symmetry.

In our case, the interaction \eqref{SU2delcurr} preserves conformal symmetry, and the fact that it does determines $\chi_\text{op}(r)$. For $d=6+\epsilon$, one can use the EFT described in section \ref{sec:closed} to calculate $\chi_\text{op}(r)$. Like there, as $\epsilon$ increases, the corrections to the EFT analysis grow, and eventually one must analyze the full boundary theory \eqref{SU2delcurr}. The resulting CFT has an enhanced $SU(2)$ symmetry for all $d$, like in the corresponding closed string analysis in section \ref{sec:closed}. 

The analogy to the closed string case suggests considering the large $d$ limit of the D-brane system described in this section. In the closed string case, the EBH solution \eqref{bh}, \eqref{fbeta} approaches in this limit the $SL(2,\mathbb{R})/U(1)$ two dimensional EBH (after reduction on the sphere) \cite{Soda:1993xc,Emparan:2013xia,Chen:2021emg}. It is interesting to do the same here, starting from the solution of \cite{Callan:1997kz}, \eqref{lphi}. 

For large $d$, the solution \eqref{lphi} has the property that $X(r)\simeq\pm\frac{L}{2}$, except in a small region around $r=r_\text{min}$. To zoom in on this region, it is convenient to define a new radial coordinate $\rho$, 
\ie
e^{\rho}=\left(\frac{r}{r_{\text{min}}}\right)^{d-1}.
\fe
The shape of the brane, \eqref{lphi}, approaches at large $d$ (and fixed $\rho$) 
\ie
\begin{split}
\label{formlarged}
X&=\pm\left(\frac{L}{2}-\frac{r_{\text{min}}}{d}\int_\rho^\infty d\rho' \frac{1}{\sqrt{e^{2\rho'}-1}}\right)\\
&=\pm\frac{r_{\text{min}}}{d}\text{arctan} \sqrt{e^{2\rho}-1}\ .
\end{split}
\fe
Using the fact that as $d\to\infty$, $r_\text{min}$ \eqref{Lrmin} approaches $\frac{d}{\pi}L$, we can rewrite \eqref{formlarged} as
\ie
\label{hpin}
e^{-\rho}=\cos \frac{Q X}{2}\ ,
\fe
with
\ie
\label{QL}
Q=\frac{2\pi}{L}~.
\fe
Further rescaling $\rho$, $\rho=Q\phi/2$, \eqref{hpin} describes a hairpin brane in a flat space labeled by $(X,\phi)$, and linear dilaton in the $\phi$ direction with slope $Q/2$~\cite{Lukyanov:2003nj}. In our application, the linear dilaton arises from the integration over the $d-1$ dimensional sphere $e^{-\Phi}\sim r^{d-1}\sim e^{\rho}=e^{Q\phi/2}$. 

Thus, we conclude that just like the Euclidean Schwarzschild solution in asymptotically flat spacetime approaches at large $d$ the two dimensional $SL(2,\mathbb{R})/U(1)$ EBH, its open string analog \eqref{lphi} approaches at large $d$ the hairpin brane, which is an open string analog of the two dimensional black hole~\cite{Kutasov:2005rr}.

Like in the EBH discussion, the DBI analysis leading to \eqref{formlarged} is only valid for small $Q$, i.e. large $L$ \eqref{QL}, but following~\cite{Chen:2021emg} one can extend the analysis to $Q\sim m_s$ and in particular study the large $d$ limit of the system with $m_\infty=0$, i.e. $L=L_c$ \eqref{llcc}. In that case, one can show that the hairpin brane preserves an $SU(2)$ global symmetry, which is consistent with our claim that this symmetry is preserved for all $d$ (and $m_\infty=0$), being a property of the worldsheet theory \eqref{SU2delcurr}.

\section{Discussion}

One of the main goals of this note was to discuss the question of what happens to the large Euclidean black hole background in string theory when the Hawking temperature of the black hole is increased to the vicinity of the Hagedorn temperature. A natural expectation is that in this limit it can be described in terms of the effective action studied by Horowitz and Polchinski \cite{Horowitz:1997jc} in the context of the string/black hole correspondence \cite{Horowitz:1996nw}. 

The original HP analysis found solutions with the right properties only in $d<6$ spatial dimensions. By generalizing their analysis to non-integer $d$, and focusing on the behavior near $d=6$, we showed (in section \ref{sec:closed}) that suitable solutions do exist for $d\ge 6$, but they require going beyond the original HP effective action \cite{Horowitz:1997jc}. For $d=6+\epsilon$, one can study these solutions in a power series in $\epsilon$ (and $m_\infty$ \eqref{mR}), by including in the effective action additional terms. One can arrange these terms by their scaling dimension w.r.t. a certain scaling symmetry. Terms of a particular scaling dimension determine the solution to a particular order in $\epsilon$. For $\epsilon$ of order one, one needs to go beyond the EFT and study the full worldsheet theory describing the EBH. 

When the Hawking temperature is {\it equal} to the Hagedorn temperature, we argued that for all $d>6$, the EBH background exhibits an enhanced $SU(2)$ symmetry. It is described by a sigma model on $\mathbb{R}^d\times S^1$, perturbed by a non-abelian Thirring deformation with a coupling that depends on the radial direction in $\mathbb{R}^d$, \eqref{JJbar}. The radial profile of the coupling is determined by the requirement that the resulting theory is conformal. 

We discussed the relation of our results to the large $d$ analysis of~\cite{Soda:1993xc,Emparan:2013xia,Chen:2021emg}, and argued that the two are compatible. In particular, the enhanced $SU(2)$ symmetry of the EBH's at the Hagedorn temperatures goes over at large $d$ to the known symmetry of the $SL(2,\mathbb{R})/U(1)$ coset CFT at level $k=4(2)$ in the (super)string. 
 
We also discussed (in section \ref{sec:open}) an open string analog of the EBH problem. The role of the black hole is played in this case by a system of two D-branes connected by a tube, whose width depends on the separation of the branes, $L$. This separation plays a similar role to that of the inverse Hawking temperature $\beta$ in the black hole analysis. 

The analog of taking the Hawking temperature to the Hagedorn temperature is in this case taking $L$ to the critical value for which a string stretched between the two branes goes to zero mass. The same logic as before motivates a description of the resulting brane configuration in terms of an effective action, which takes a similar form to the closed string one. 

We showed that the logic used in section \ref{sec:closed} for the EBH case leads to a sensible picture in this case as well. In particular, when the separation of the branes $L$ is equal to the critical one, the system again exhibits an enhanced $SU(2)$ symmetry, and is described by a kind of Kondo Lagrangian with an $r$-dependent coupling, \eqref{SU2delcurr}. 

The analogy of the closed and open string systems suggests that at large $d$ the D-brane solution should approach (at any $L$) the open string analog of the $SL(2,\mathbb{R})/U(1)$ coset CFT, the hairpin brane of \cite{Lukyanov:2003nj}. We showed that this is indeed the case. At the critical value of $L$, the hairpin has an $SU(2)$ symmetry, that can be viewed as the large $d$ limit of the $SU(2)$ symmetry of \eqref{SU2delcurr}.

The main qualitative conclusion from our results is that the hypothesis that HP-type solutions correspond to the continuation of EBH's to the vicinity of the Hagedorn temperature is consistent with all the tests we subjected it to. 

As mentioned in section \ref{sec:intro}, the authors of \cite{Chen:2021dsw} argued that in the case of $(1,1)$ worldsheet supersymmetric models, solutions of the sort we studied in section \ref{sec:closed} differ from Euclidean Schwarzschild black holes in their Witten indices and spectrum of D-branes. One can use our improved understanding of these models to further study these issues.

There are a number of natural extensions of this work that would be interesting to explore. Our analysis gave rise to some interesting new CFT's, such as the non-abelian Thirring model with an $r$-dependent coupling \eqref{JJbar}, and the Kondo-type boundary CFT \eqref{SU2delcurr}. It would be nice to solve these CFT's and study their properties. This might elucidate the relation of these theories to their large $d$ limits (the cigar and hairpin, respectively).  

Another interesting direction is to study the Lorentzian analogs of the Euclidean backgrounds discussed in this note. The Lorentzian analog of the background of section \ref{sec:closed} is the Lorentzian black hole. Studying it at temperatures near the Hagedorn temperature is an interesting way to analyze the stringy effects on its horizon, and the string/black hole transition of \cite{Horowitz:1996nw}. 

A concrete challenge is to understand the condensate of the winding tachyon $\chi$ from a Lorentzian point of view. This was discussed already in the original paper \cite{Horowitz:1997jc}, and for the related $SL(2,\mathbb{R})/U(1)$ black hole in some more recent work, e.g. \cite{Kutasov:2005rr,Giveon:2020xxh,Jafferis:2021ywg} and references therein. However, the problem remains largely unsolved, since in this case one does need to address the region beyond the horizon and the black hole singularity. Solving it might contribute to our understanding of the microstates of small Schwarzschild black holes in string theory.  

The analogy of the open and closed string systems makes it interesting to consider the Lorentzian continuation of the Euclidean background of section \ref{sec:open}. In the case of the two dimensional black hole and hairpin, this was discussed in \cite{Kutasov:2005rr}, where it was pointed out that the Lorentzian analog of the hairpin is an accelerating $D0$-brane in a spacelike linear dilaton background, and that the stringy effects associated with the stretched tachyon  $\chi_\text{op}$ correspond to a stringy smearing of the trajectory of the $D0$-brane, that becomes more pronounced as its Unruh temperature approaches the Hagedorn temperature. 

In the flat spacetime system discussed in section \ref{sec:open} of this note, the Lorentzian continuation describes a spherical D-brane in flat spacetime, that expands to a maximal size, and then shrinks back. Finding a solution of the corresponding boundary CFT would be interesting for analyzing the stringy effects for this brane, and might improve our understanding of the analogous closed string system (the black hole). 

In both the closed and open string systems, we found that the case of $d=6$ space dimensions plays a special role. In particular, the nature of the black hole solutions near the Hagedorn temperature is different for $d<6$ and $d>6$. It would be interesting to understand better the origin of the difference, and the implications for the string/black hole transition.

\section*{Acknowledgements}

This work was supported in part by DOE grant DE-SC0009924 and BSF grant 2018068. The work of DK was also supported in part by National Science Foundation grant PHY-1607611 at the Aspen Center for Physics. DK thanks the Weizmann Institute, Tel Aviv University, and the Aspen Center for Physics for hospitality during part of this work.

\appendix

\section{Scaling analysis of the effective action}
\label{app:EFTscaling}
In this appendix, we derive a relation (which is used in the text) between different parts of the action \eqref{actioncl} evaluated on classical solutions, that satisfy suitable boundary conditions.

Let $(\chi_*(x),\varphi_*(x))$ be a classical solution of \eqref{actioncl}. Performing a rescaling of the fields: 
\begin{equation}
\label{chiphiscaling}
    \begin{split}
        &\chi(x)=\lambda_\chi \chi_*\left(\frac{x}{\gamma}\right)~,\\
        &\varphi(x)=\lambda_\varphi \varphi_*\left(\frac{x}{\gamma}\right)
    \end{split}
\end{equation}
leads to the action
	\begin{equation}
	\label{Sphichirescaled}
	\begin{split}
 	I_d(\lambda_\chi,\lambda_\varphi,\gamma)=&\frac{1}{16\pi G_N}\int d^dx\bigg[\gamma^{d-2}\lambda^2_\varphi(\nabla\varphi_*)^2+\gamma^{d-2}\lambda_\chi^2|\nabla \chi_*|^2\\
 	&+\gamma^d\left(\lambda_\chi^2m^2_\infty+\lambda_\varphi\lambda_\chi^2\frac{\kappa}{\alpha'}\varphi_*+\lambda_\varphi^2\lambda_\chi^2\frac{\kappa}{\alpha'}\varphi_*^2\right)|\chi_*|^2+\gamma^d\lambda_\chi^4\frac{\kappa}{4\alpha'}|\chi_*|^4\bigg]\ .
	\end{split}
	\end{equation}
Stationarity of the action on the classical solution $(\chi_*,\varphi_*)$ implies that 
\begin{equation}
\label{saddle}
    \begin{split}
         \left.\frac{\partial I_d}{\partial \lambda_\chi}\right|_{\lambda_\chi=\lambda_\varphi=\gamma=1}=\left.\frac{\partial I_d}{\partial \lambda_\varphi}\right|_{\lambda_\chi=\lambda_\varphi=\gamma=1}=\left.\frac{\partial I_d}{\partial \gamma}\right|_{\lambda_\chi=\lambda_\varphi=\gamma=1}=0\ .
    \end{split}
\end{equation}
One of the three resulting relations is
\begin{equation}
\label{mchiint}
    \begin{split}
      m_{\infty}^2\int d^d x  |\chi_*|^2=\frac{1}{4}\int d^dx\left[(d-6)\frac{\kappa}{\alpha'}\varphi_*|\chi_*|^2+(2d-8)\left(\frac{\kappa}{\alpha'}\varphi_*^2|\chi_*|^2+\frac{\kappa}{4\alpha'}|\chi_*|^4\right)\right]\ .
    \end{split}
\end{equation}
In deriving \eqref{mchiint} we assumed that the different integrals on the l.h.s. and r.h.s. are finite. In this note we are interested in sperically symmetric solutions that satisfy the boundary conditions \eqref{largeRbc} at large $r$ and go to constants as $r\to 0$, for which this is indeed the case.

\section{Derivation of the open string effective action}
\label{Smat}

In this appendix we derive the effective action \eqref{opeft}, by studying the open string S-matrix. We do the calculation in the bosonic string, but expect the structure to be similar in the type II case. 

Our main focus is on the coefficients of the quartic terms in $\chi_\text{op}$ and $\varphi_\text{op}$ in \eqref{opeft}. We do the calculation at $L=L_c$; corrections in $(L-L_c)/L_c$ give higher order contributions in $m_\infty$, \eqref{moL}, that are subleading in the expansion described in the text.

The worldsheet CFT is defined in this case on the upper half plane. The boundary conditions correspond to open strings ending on one of the two D-branes described in section \ref{sec:open}. Thus, these strings have Chan-Paton factors described by $2\times 2$ matrices. 

The spacetime field $\varphi_\text{op}$ corresponds to a string with both ends on the same D-brane; its Chan-Paton factor corresponds to the Pauli matrix $\sigma^3$. $\chi_\text{op}$ describes a string going from one D-brane to the other. Its Chan-Paton factor corresponds to $\sigma^-\equiv \frac{1}{2}\left(\sigma^1-i \sigma^2\right)$. 

At $L=L_c$, the boundary vertex operator of $\chi_\text{op}$ is proportional to 
\begin{equation}
\label{Jplus}
	\begin{split}
 	J^+=e^{iL_c(X_L-X_R)/(2\pi\alpha')}=e^{i(X_L-X_R)/\sqrt{\alpha'}}\ ,
	\end{split}
	\end{equation}
and that of $\varphi_\text{op}$ to
\begin{equation}
\label{J3}
	\begin{split}
 	J^3=\frac{1}{2\sqrt{\alpha'}}\partial_nX=\frac{i}{2\sqrt{\alpha'}}(\partial X_L-\bar{\partial}X_R)\ .
	\end{split}
	\end{equation}	
The OPE of $J^1=(J^++J^-)/2$, $J^2=(J^+-J^-)/2i$ and $J^3$ is (here $z\in\mathbb{R}$ parametrizes the boundary of the upper half plane)
\begin{equation}
	\begin{split}
 	J^i(z)J^j(0)\sim \frac{\delta^{ij}}{2z^2}+i\frac{\epsilon^{ijk}}{z}J^k(0)\ .
	\end{split}
	\end{equation}
To calculate the effective action of $\chi_\text{op}$ and $\varphi_\text{op}$, we need to evaluate the S-matrix elements of the corresponding vertex operators,
$a_\chi (r)$, $a_{\chi^*}(r)$ and $a_\varphi (r)$. As discussed above, they are given by 
\ie
\label{aphichi}
\begin{split}
e^{ik\cdot r}J^+\sigma^-\ ,\\
e^{ik\cdot r}J^-\sigma^+\ ,\\
e^{ik\cdot r}J^3\sigma^3\ .
\end{split}
\fe

We start with the relative normalization of the kinetic terms in the effective action, 
\ie
G_{ij}\nabla \phi^i\cdot \nabla\phi^j\ .
\fe
The coefficient $G_{ij}$ is proportional to the two-point function of the vertex operators for the fields $\phi^i$ and $\phi^j$:
\ie
\label{2ptfunc}
\begin{split}
G_{\chi\chi^*}&\sim\langle e^{ik\cdot r}J^+(1)e^{-ik\cdot r}J^-(0)\rangle \text{Tr}\sigma^-\sigma^+=1\ ,\\
G_{\varphi\varphi}&\sim\langle e^{ik\cdot r}J^3(1) e^{-ik\cdot r}J^3(0)\rangle\text{Tr}\sigma^3\sigma^3=1\ .
\end{split}
\fe
Therefore, the coefficient of $|\nabla a_\chi|^2$ is twice as large as that of $(\nabla a_\varphi)^2$. 

The scattering amplitude for $a_\chi a_\chi^*a_\varphi$ is given by
\ie
\label{chi2phi}
\begin{split}
S_{\chi\chi^*\varphi}{(k_1;k_2;k_3)}=&\text{Tr}(\sigma^-\sigma^+\sigma^3)\left\langle e^{ik_1\cdot r(\infty)}J^+(\infty)e^{ik_2\cdot r(1)}J^-(1)e^{ik_3\cdot r(0)}J^3(0)\right\rangle\langle c(\infty)c(1)c(0)\rangle\\
&+\text{Tr}(\sigma^-\sigma^3\sigma^+)\left\langle e^{ik_1\cdot r(\infty)}J^+(\infty)e^{ik_3\cdot r(1)}J^3(1)e^{ik_2\cdot r(0)}J^-(0)\right\rangle\langle c(\infty)c(1)c(0)\rangle\\
=&-2 C_{D_2}(2\pi)^d\delta^d(k_1+k_2+k_3)\ .
\end{split}
\fe
Here $C_{D_2}$ is a constant discussed e.g. in \cite{Polchinski:1998rq}. The cubic coupling $a_\chi a_\chi^*a_\varphi$ in the effective action can be obtained by comparing to the scattering amplitude \eqref{chi2phi}.

The quartic terms in the effective action can be obtained by studying $2\to 2$ scattering amplitudes of \eqref{aphichi}. 

For four $a_\varphi$ we have: 
\ie
\label{phi4}
\begin{split}
S&_{\varphi\varphi\varphi\varphi}{(k_1;k_2;k_3;k_4)}=\int_0^1 dx\left\langle e^{ik_1\cdot r(\infty)}J^3(\infty)e^{ik_2\cdot r(1)}J^3(1)e^{ik_3\cdot r(x)}J^3(x)e^{ik_4\cdot r(0)}J^3(0)\right\rangle\\
&\text{Tr}(\sigma^3\sigma^3\sigma^3\sigma^3)\langle c(\infty)c(1)c(0)\rangle+\text{permutations of }(k_2,k_3,k_4)\\
=&C_{D_2}(2\pi)^d\delta^d(k_1+k_2+k_3+k_4)\bigg[2\int_0^1(1-x)^{2\alpha'k_2\cdot k_3}x^{2\alpha'k_3\cdot k_4}\left(\frac{1}{4x^2}+\frac{1}{4(1-x)^2}+\frac{1}{4}\right)\\
&+\text{permutations of }(k_2,k_3,k_4)\bigg]\\
=&C_{D_2}(2\pi)^d\delta^d(k_1+k_2+k_3+k_4)\frac{2\alpha'^2\pi^2}{3}\bigg[(k_2\cdot k_3)(k_3\cdot k_4)+2(k_2\cdot k_3)^2\\
&+\text{cyclic permutations of }(k_2,k_3,k_4)\bigg]+O(k^6)\ .
\end{split}
\fe
As is familiar, this is analytic in Mandelstam invariants, since $\varphi$ does not have cubic couplings of the form $\varphi^2O$, with any massless open string field $O$. The terms quartic in momenta in \eqref{phi4} follow from the expansion of the DBI action for $\varphi$ to quartic order. 

For two $a_\chi$ and two $a_{\chi^*}$ we find
\ie
\label{chi4}
\begin{split}
S&_{\chi\chi^*\chi\chi^*}{(k_1;k_2;k_3;k_4)}=\int_0^1 dx\left\langle e^{ik_1\cdot r(\infty)}J^+(\infty)e^{ik_2\cdot r(1)}J^-(1)e^{ik_3\cdot r(x)}J^+(x)e^{ik_4\cdot r(0)}J^-(0)\right\rangle\\
&\text{Tr}(\sigma^-\sigma^+\sigma^-\sigma^+)\langle c(\infty)c(1)c(0)\rangle+(k_2\leftrightarrow k_4)\\
=&C_{D_2}(2\pi)^d\delta^d(k_1+k_2+k_3+k_4) 2\int_0^1(1-x)^{2\alpha'k_2\cdot k_3-2}x^{2\alpha'k_3\cdot k_4-2}\\
=&2C_{D_2}(2\pi)^d\delta^d(k_1+k_2+k_3+k_4)\\
&\bigg[\frac{2-\alpha'k_2\cdot k_3+\alpha'k_2\cdot k_4}{2\alpha'k_3\cdot k_4}+\frac{2-\alpha'k_3\cdot k_4+\alpha'k_2\cdot k_4}{2\alpha'k_2\cdot k_3}-1+2\left(2+\frac{\pi^2}{3}\right)k_2\cdot k_4+O(k^4)\bigg]\ .
\end{split}
\fe
The first two terms in the last expression come from the exchange of $\varphi$ and the gauge field on the branes, $A_i$. As discussed in \cite{Polchinski:1998rq}, one can use them to compute $C_{D_2}$. For example, the $\varphi$ exchange in the first term gives
\ie
\begin{split}
\frac{2C_{D_2}}{\alpha'k_3\cdot k_4}(2\pi)^d\delta^d(k_1+k_2+k_3+k_4)&=\int\frac{d^dk}{(2\pi)^d}\frac{2G_{\varphi\varphi}S_{\chi\chi^*\varphi}(k_1;k_2;k)S_{\varphi\chi\chi^*}(-k;k_3;k_4)}{k^2}\\
&=\frac{C_{D_2}^2}{G_{\varphi\varphi}k_3\cdot k_4}(2\pi)^d\delta^d(k_1+k_2+k_3+k_4)\ .
\end{split}
\label{CD2unitarity}
\fe
Therefore,
\ie
C_{D_2}=\frac{2G_{\varphi\varphi}}{\alpha'}\ .
\label{CD2}
\fe
After subtracting it, we find a constant term, that comes from the interaction Lagrangian
\ie
\frac{G_{\varphi\varphi}}{\alpha'}|a_\chi|^4\ .
\fe

For two $a_\varphi$, one $a_\chi$, and one $a_{\chi^*}$, we have
\ie
\label{chi2phi2}
\begin{split}
S_{\chi\chi^*\varphi\varphi}(k_1;&k_2;k_3;k_4)=\int_0^1 dx\left\langle e^{ik_1\cdot r(\infty)}J^+(\infty)e^{ik_2\cdot r(1)}J^-(1)e^{ik_3\cdot r(x)}J^3(x)e^{ik_4\cdot r(0)}J^3(0)\right\rangle\\
&\times\text{Tr}(\sigma^-\sigma^+\sigma^3\sigma^3)\langle c(\infty)c(1)c(0)\rangle\\
&+\int_0^1 dx\left\langle e^{ik_1\cdot r(\infty)}J^+(\infty)e^{ik_3\cdot r(x)}J^3(1)e^{ik_2\cdot r(1)}J^-(x)e^{ik_4\cdot r(0)}J^3(0)\right\rangle\\
&\times\text{Tr}(\sigma^-\sigma^3\sigma^+\sigma^3)\langle c(\infty)c(1)c(0)\rangle\\
&+\int_0^1 dx\left\langle e^{ik_1\cdot r(\infty)}J^+(\infty)e^{ik_3\cdot r(1)}J^3(1)e^{ik_4\cdot r(x)}J^3(x)e^{ik_2\cdot r(1)}J^-(0)\right\rangle\\
&\times\text{Tr}(\sigma^-\sigma^3\sigma^3\sigma^+)\langle c(\infty)c(1)c(0)\rangle+(k_3\leftrightarrow k_4)\\
=&C_{D_2}(2\pi)^d\delta^d(k_1+k_2+k_3+k_4)\bigg[\int_0^1(1-x)^{2\alpha'k_2\cdot k_3-1}x^{2\alpha'k_3\cdot k_4}\\
&+\frac{1}{2}\int_0^1(1-x)^{2\alpha'k_2\cdot k_3}x^{2\alpha'k_3\cdot k_4-2}+\int_0^1(1-x)^{2\alpha'k_2\cdot k_3-1}x^{2\alpha'k_2\cdot k_4-1}\\
&-\frac{1}{2}\int_0^1(1-x)^{2\alpha'k_2\cdot k_3}x^{2\alpha'k_2\cdot k_4}+\int_0^1(1-x)^{2\alpha'k_3\cdot k_4}x^{2\alpha'k_2\cdot k_4-1}\\
&+\frac{1}{2}\int_0^1(1-x)^{2\alpha'k_3\cdot k_4-2}x^{2\alpha'k_2\cdot k_4}+(k_3\leftrightarrow k_4)\bigg]\\
=&C_{D_2}(2\pi)^d\delta^d(k_1+k_2+k_3+k_4)\\
&\bigg[\frac{2}{\alpha'k_2\cdot k_3}+\frac{2}{\alpha'k_2\cdot k_4}-2-4\alpha'\left(1+\frac{\pi^2}{6}\right)k_3\cdot k_4+O(k^4)\bigg] .
\end{split}
\fe
The first two terms in the last expression are again due to exchange diagrams. 
After subtracting them, we are left with the leading interaction term
\ie
\frac{G_{\varphi\varphi}}{\alpha'}a_\varphi^2|a_\chi|^2\ .
\fe

The normalization factor $G_{\varphi\varphi}$ can be determined by demanding that the EFT has the same normalization for the kinetic term $(\nabla X)^2$ as the DBI action \eqref{dbiact}. To relate $X$ to $a_\varphi$, we note that the $a_\varphi|a_\chi|^2$ term in the action can be understood as the expansion of the mass of the tachyon, \eqref{moL}, with
\ie
\label{Lx}
L(x)=|2X(x)|= L\left(1-a_\varphi(x)\right)+{\cal O}(a_\varphi^2)\ .
\fe
Therefore, we have $\frac{\tau_d}{2}(\nabla X)^2= \frac{\tau_dL^2}{8}(\nabla a_\varphi)^2+{\cal O}(a_\varphi^3)$, where we read off that
\ie
G_{\varphi\varphi}=\frac{\tau_dL^2}{8}\ .
\fe

Combining all the above results, we arrive at the effective action
	\begin{equation}
	\label{modeft}
	\begin{split}
I_\text{op}= \frac{\tau_dL^2}{8}\int d^dx\left[(\nabla a_\varphi)^2+2|\nabla a_\chi|^2+2\left(m_\infty^2-\frac{2}{\alpha'}a_\varphi+\frac{1}{\alpha'}a_\varphi^2\right)|a_\chi|^2+\frac{1}{\alpha'}|a_\chi|^4\right]\ .
	\end{split}
	\end{equation}
As is clear from the derivation, this action receives contributions of higher order in fields and derivatives. 	
Redefining $a_\chi\to \sqrt{2}\chi_\text{op},a_\varphi\to -2\varphi_\text{op} $, we get the action \eqref{opeft}.

\vskip 2cm

\bibliographystyle{JHEP}
\bibliography{HP2}
%%%%%%%%%%%%%%%%%%%%%%%%%%%%%%%%%%%%%%
%%%%%%%%%%%%%%%%%%%%%%%%%%%%%%%%%%%%%%

\end{document}